\documentstyle[12pt]{article}
\def\ZZ{\hbox{$Z$}
        \hbox{\kern -6pt  Z}}
\begin{document}
\pagestyle{empty}
\rightline{LPT Strasbourg 96-07}\par
\rightline{hep-th/9603149}\par
\begin{center}
 {\large \bf   $2D-$ Fractional Supersymmetry for Rational Conformal Field
 Theory: Application for Third-Integer Spin  States  } 
\end{center}
 
 \vskip 1 truecm

\begin{center}
    {\large A. Perez}$^1$ \\
    {\large M. Rausch de Traubenberg}$^{1,2}$ \\
    and \\
    {\large P. Simon}$^1$

$^1${\large Laboratoire de Physique Th\'eorique, Universit\'e Louis Pasteur} \\
  {\large 3-5 rue de l'universit\'e, 67084 Strasbourg Cedex, France} \\
$^2$  
  {\large Centre de Recherches Nucl\'eaires, B\^at 40/II 67037 Strasbourg 
  Cedex 2 }\\  
\end{center}

\vskip 1 truecm
\abstract{A $2D-$ fractional supersymmetry  theory is algebraically
constructed. The Lagrangian is derived using an adapted superspace 
including, in addition to a scalar  field, two fields with spins $1/3,2/3$.
This theory turns out to be a rational conformal field theory. 
The symmetry of this model goes beyond the super Virasoro algebra 
and connects these third-integer spin states. Besides
the stress-momentum tensor, we obtain a supercurrent of spin $4/3$.
Cubic  relations are involved in order to close the algebra;
the basic algebra is no longer a Lie or a super-Lie algebra. The central charge 
of this model is  found to be $5/3$. Finally, we analyse the form that a local
invariant action should take.} 
\vskip 2truecm
\centerline{ Nucl. Phys. {\bf B 482} (1996) 325.} 
\vfill
\eject
 \pagestyle{plain}
\pagenumbering{arabic} 
\vskip .5 truecm  
\noindent
{\bf  I. Introduction.}\\

\vskip .5 truecm
$2D-$ conformal invariance, after the work of Belavin, Polyakov and 
Zamolodchikov \cite{bpz}, becomes a formidable tool for the description
of $2D-$ critical phenomena and   string theory. In that context, a study of
$2D$ conformal field constitutes a great challenge for a 
classification of integrable models, and    a description of  $4D-$string 
(in order to obtain string solutions with a good phenomenology).
The first attempt, in a systematic  classification,
has been done by Friedan, Qiu and Shenker \cite{fqs1} who argued that if
one imposes unitarity and a spectrum bounded from below ( highest weight 
representation) one  gets, within the framework of the Virasoro algebra,
constraints on the values of the central charges $c$ and the conformal weights
$h$. They have obtained two different kinds of integrable models:
one  with $c>1$ (and with an infinite number of primary fields), and   discrete
series for $c<1$. However, it has been proved that if we enlarge the symmetry 
of the $2D-$manifold, other series do appear. For instance, with a $N=1$
superconformal algebra, other integrable models can be described with 
$c < 3/2$ \cite{fqs2}. Meanwhile, if one extends the
symmetry of the worldsheet of the string, in the
framework of super Virasoro algebra, the critical dimension  
goes consequently  from  $26$   to $10$ \cite{ps}. If one   takes a  $N=2$
super-Virasoro algebra,  as the basic symmetry of the worldsheet, the critical
dimension is then $D=2$.

Nevertheless, the Virasoro algebra can be extended using the parafermions
introduced by Fateev and Zamolodchikov \cite{fz}, and leading to new
series of exact solvable models \cite{fconf}. Noting that those theories
naturally contain   a  current of spin  $K+4 \over K+2$  ($K=2$ corresponds
to the superconformal algebra), they have been applied in the context of
string theory with a critical dimension $2 + {16 \over K}$ \cite{t}.
All those solutions have a common feature, they can be realized in terms
of the  coset construction of Goddard, Kent and Olive (GKO) \cite{gko} with 
appropriate Kac-Moody  algebras \cite{go}. Let us point out that the GKO
construction can be applied with all kinds of affine Lie algebras. For 
instance, the $W_n$ algebras \cite{w}, obtained in this approach,
involving primary fields up to spin $n$, close in its universal
enveloping algebra; and so cannot be defined as a Lie algebra.

In the present paper, we will follow another direction to extend the Virasoro
algebra by introducing currents of fractional spin, which does
not close through quadratic relations, as $W_n$ algebras.
Our starting point is, neither
the parafermions, nor the GKO coset construction, but the interesting property
of $1,2,3 ~D-$~spaces where the states are not in a representation of
the permutation group but rather of the braid group \cite{lm}. 
(We can point out that the parafermions have also non trivial monodromy
transformations). This situation, previously exploited in order to
extend supersymmetry to fractional supersymmetry
\cite{fsusy1d,fsusy2d,am,fr},  has    been considered in $1D$
\cite{fsusy1d,am,fr} where this symmetry can be seen as a $  F^{  th}$
root of the time translation $\partial_t$; or in
$2D$   as a $   F^{  th}$
root of conformal transformations. $F=2$ corresponds to
the usual supersymmetry.
This procedure, has been already applied
in the case of $1D$ fractional supersymmetry. It leads to a new
equation  acting on the states which are in the representation of
the braid  group \cite{fr}. The method adopted there is similar 
to the one leading to the Dirac equation in $1D$ using the
supersymmetry \cite{susy1d}.

Here, we study $2D-$ fractional supersymmetry, {\it i.e.} we
extend the Virasoro algebra with a current of spin ${1+F \over F}$. In
addition to the scalar field we introduce fields of conformal weight
${1 \over F} \dots {F-1 \over F}$. It turns out that the fractional 
supersymmetry is a symmetry which connects states of fractional spin.
For $F=3$, the central charge is found to be
$c=5/3$, proving that this construction
is different from the one of Ref. \cite{fconf}, because the central charge
is, $c={3K \over K+2}$ ($c=2$ for $K=4$).  When $F=3$, we have a conserved
current of spin $4/3$ like in the case $K=4$.
So, the conformal weights
of this construction cannot be obtained through the approaches detailed in
Ref. \cite{fconf}. As far as we know, no GKO approach of this model
has been built up.

In the literature, a similar approach has been already obtained by Saidi
{\it et al.} \cite{ssz}. These authors have also introduced fractional spin
for fractional supersymmetry using parafermions and involving non
local Operator Product Expansion (O.P.E.). A more detailed analysis
of this point will be given in Section IV.

This paper will be divided as follow:  Sect. II is devoted to 
a description of the main results already obtained in $1D$. In Sect. III, 
we construct explicitly the $2D$ fractional supersymmetric lagrangian, 
introducing an adapted fractional superspace by help of Generalized Grassmann 
variables
and its differential structure. In Sect. IV, we calculate the Green function
and we propose a normal ordering prescription in order to apply of the Wick
theorem. We also discuss the $q-$mutation relations of the modes  
of the fields. In Sect. V, we determine the algebraic structure of the basic
fields on behalf of the OPE. Sect. VI. is devoted to build action beyond
FSUSY, {\it i.e.} by gauging the global symmetries.
Finally, in the conclusion, we give an outlook of these new algebras, 
and obtain new  critical dimensions for string.

\vskip .5 truecm  
\noindent
{\bf  II. Summary of the Main Results of $1D-$ Fractional Supersymmetry.}\\

\vskip .5 truecm 
 
Supersymmetry, which is the only non-trivial $Z_2$ extension of the Poincar\'e
algebra \cite{cm,hls}, can be
naturally generalized to fractional supersymmetry
\cite{fsusy1d,fsusy2d,am,fr}, 
as soon as the space-time dimension is smaller than $3$, where
alternative statistics are allowed. 
The algebraic structure of fractional supersymmetry (FSUSY), possesses a
$Z_F$ structure (here, in this paper, we will consider only the case $F=3$ )
so the basic fields will be of graduation $0,\dots F-1$, generalizing
the concept of boson/fermion which are respectively of graduation $0,1$
{\it i.e} even/odd with respect to $Z_2$.

\noindent
Let us recall the main results of this symmetry in one dimension
(this symmetry has already been introduced in $1D$ 
as a global symmetry \cite{fsusy1d,am} or as a local one \cite{fr}).

\noindent
FSUSY is generated by $H$, the Hamiltonian 
(or the generator of time translation) and $Q$, the generator of the FSUSY
transformations. The algebra fulfills 
\begin{eqnarray}
[~Q,~H]&=&0\\ 
Q^{3}&=&-H, \nonumber
\label{1.1}
\end{eqnarray}

\noindent
it is important to emphasize that the algebra (1) is neither a 
Lie algebra
nor a superalgebra, because it closes through a cubic relation, and goes beyond
the framework of the Coleman, Mandula \cite{cm} and Haag, Sohnius, 
Lopuszanski \cite{hls} theorems, which deal with Lie or super Lie algebras.
It is interesting to notice that most results of supersymmetry and
supergravity ( see {\it e.g.} \cite{wb}) can be transposed easily within
the framework of fractional supersymmetry  (for more details see 
\cite{fsusy1d,fsusy2d,am,fr}).
This symmetry acts on an analogous of a superspace introduced in supersymmetry
(SUSY). The time $t$ is then  extented to $(t, \theta)$ with $\theta$ 
a real generalized  Grassmann variable  ($\theta^3=0$) \cite{gca1}, instead
of a Grassmann one.
The introduction of $\epsilon$ and $f$, the parameters of the FSUSY 
transformations
and the time translation, leads to the transformations \cite{am,fr} 
\begin{eqnarray}
t'&=&t+q (\epsilon^2 \theta +\epsilon \theta^2 )-f\\
{\theta} '&=&\theta+\epsilon,\nonumber
\label{1.2} 
\end{eqnarray}

\noindent 
$\epsilon$  verifies $\epsilon^3=0$ and $\theta {\epsilon}=q \epsilon \theta$,
with $q=\exp{({2 i \pi \over 3})}$. 
The $q-$ mutation between the two variables  $\epsilon$ and $\theta$ has 
four origins :

$\bullet$ it ensures that if $\epsilon^3=\theta^3=0$ then $(\epsilon + \theta)^3
=0$ \cite{gca1};

$\bullet$  the time remains real after a FSUSY transformation;

$\bullet$ the FSUSY transformations commute  with the covariant derivative
(see after);

$\bullet$  the FSUSY  transformations ($\epsilon Q$, see
after) satisfy the Leibnitz rule [see Durand  in \cite{fsusy1d}].

\noindent
Next, we consider a real fractional superfield $\Phi$ 
in the scalar representation
of the fractional superline  

\begin{equation}
\Phi(t, \theta) = x(t) + q^2 \theta \psi_2(t) + q^2 \theta^2 \psi_1(t),
\label{1.3}
\end{equation} 

\noindent
where $x(t), \psi_1(t), \psi_2(t)$ are the extensions of the bosonic and
fermionic fields. They satisfy   
$\psi_1^3=\psi_2^3=0$, 
and  their grade is respectively $0$,$1$ and  $2$ . 
 They  are submitted to the
$q-$mutation relations (postulated from their grade) 
\begin{eqnarray}
\theta \psi_1(t)  &=& q  \psi_1(t) \theta \nonumber \\
\theta \psi_2(t)  &=& q^2  \psi_2(t) \theta \\
\psi_2(t) \psi_1(t) &=& q \psi_1(t) \psi_2(t), \nonumber
\label{1.4}
\end{eqnarray}
it can be stressed that these relations are the only ones which are arbitrary,
all the other follow naturally \cite{fr}. 
Using relations (2), we get easily the FSUSY transformations
upon the fields 
\begin{eqnarray}
\delta_{\epsilon} x &=& q^2 \epsilon \psi_2 \nonumber \\
\delta_{\epsilon} \psi_2&=& -q \epsilon  \psi_1 \\
\delta_{\epsilon} \psi_1 &=& \epsilon \dot x. \nonumber
\label{1.5}
\end{eqnarray}

\noindent
To build the action,
we   need to recall some basic features on the derivation acting
on generalized Grassmann variables. This structure,
the $q-$ deformed Heisenberg algebra, has been analyzed  in \cite{gca1}
as well as its matrix representation \cite{gca1,hq}. It  admits in general
$(F-1)$ derivatives. In our particular case,  the two derivatives are noted 
$\partial_\theta$ and $\delta_\theta$ with the properties  

\begin{eqnarray}
&&\partial_\theta \theta - q \theta \partial_\theta =  1 \nonumber \\
&&\delta_\theta \theta -  q^2 \theta \delta_\theta   = 1 \nonumber \\
&&\partial_\theta^3=0 \  \ \ \  \delta_\theta^3=0 \\
&&\partial_\theta \delta_\theta = q^2 \delta_\theta \partial_\theta.   \nonumber
\label{1.6}
\end{eqnarray}

 \noindent
Then, let us consider the two basic objets $Q$ and $D$,
which represent  respectively the FSUSY generator  
and the covariant derivative  \cite{fsusy1d,fsusy2d,am,fr}

\begin{eqnarray}
Q &=& \partial_\theta + q \theta^2 \partial_t \nonumber \\
D &=& \delta_\theta + q^2 \theta^2 \partial_t. 
\label{1.7}
\end{eqnarray}

\noindent
It can be checked explicitly that $D^3=Q^3=-\partial_t$ and
$QD=q^2 DQ$, and a direct calculation (using $\theta \epsilon
= q \epsilon \theta$)
proves that

\begin{eqnarray}
\delta_\epsilon \Phi & =& \epsilon Q \Phi(t,\theta).  
\label{1.8}
\end{eqnarray}

\noindent
Using the fact that $D$ $q-$mutes with $Q$ we have 
$\delta_\epsilon D \Phi =
D \delta_\epsilon \Phi$. Finally, arguing that the $\theta^2$ component  of
$\Phi$ transforms like a total derivative, we can construct the action   by
taking the $\theta^2$ part of the action built in the fractional superspace.
In  other words,
using the results on integration upon generalized Grassmann 
variables \cite{intgca}
$\int d\theta = {d^{n-1} \over d\theta^{n-1}}$
we obtain for $n=2$

\begin{eqnarray}
S &=& -{q^2 \over 2}\int  dt d \theta \dot \Phi D \Phi \nonumber \\
  &=& \int dt (\dot {x^2  \over 2} + {q^2 \over 2} \dot \psi_1 \psi_2
  -{q \over 2} \dot \psi_2 \psi_1). 
\label{1.9}
\end{eqnarray}

\noindent
This action has been extended under a local invariant form, introducing two
gauge fields, and leads, after quantization, to an equation  generalizing
the Dirac one. A formulation,
invariant under general reparametrization, has been given by means of a 
curved fractional superline and a analogous of a superdeterminant \cite{fr}.
 
\vskip .5truecm
\noindent
{\bf  III.   $2D-$ Fractional Supersymmetry on Riemann surfaces.}\\
\vskip .5truecm 
Now we want to extend all those results to build an action in the complex
plane ($2D$ FSUSY was introduced in 
Ref. \cite{fsusy2d}).
This $2D-$ space might be used for the description of some $2D-$
integrable models; or even should represent the symmetry of
the world-sheet of some
string theories. The first step is to construct different sets of generalized
Grassmann algebra (GGA) with its associated differential structure.

It is crucial to note that to endow the GGA  with a complex structure 
(two generalized Grassmann variables $\theta, {\bar {\theta}}$ with 
${\bar{\theta}}=\theta^{\star}$ ) is {\it clearly} incompatible with 
the $q$-mutation 
($ \theta {\bar{\theta}}=q {\bar{\theta}} \theta$)\footnote{If we would assume
$\theta \bar \theta = q \bar \theta \theta$, with  $\theta$ and $\bar \theta$
two complex conjugated variables, we get $\theta \bar \theta^2 = q^2 
\bar \theta^2 \theta$ on one hand. If we conjugate this equation we get
$\theta^2 \bar \theta = q \bar \theta \theta^2$ on the other hand. This last
equation clearly contradicts the hypothesis.}. So we cannot, as it could 
have been expected 
at first sight, generalize
the $1D$ case directly  by introducing a complex generalized Grassmann
variable. So, we have to consider an alternative construction.

Like in heterotic string \cite{gsw}, where $z$ and $\bar z$ are extented 
differently ($z \rightarrow (z,\theta)$ and $\bar z$ remains unaffected),
here, we associate to $z$ and $\bar{z}$
two {\it real} generalized Grassmann variables 
$\theta_{L}$ and $\theta_{R}$. In other words, the construction acts
separately onto the $L-$movers and $R-$movers.

Consider the generalized Grassmann variables $\theta_i$ and its two
derivatives (see (6)) $\partial_i$ and $\delta_i$ ($i$ running 
from $1$ to $p$). In the previous case, we had $p= 2$
and $\theta_1=\theta_L$, $\theta_2=\theta_R$. From
\begin{equation}
\theta_i \theta_j=q \theta_j \theta_i,~~~i < j~,
\label{2.1}
\end{equation}
\noindent
the consistency of the algebra leads to the following relations
\begin{eqnarray}
\partial_{i}\partial_j&=&q\partial_{j}\partial_i,~~~i < j\nonumber\\
\partial_i \theta_j&=& q^{-1} \theta_j \partial_i,~~~i< j\\
\partial_j \theta_i&=& q \theta_i \partial_j,~~~i< j.\nonumber
\label{2.2}
\end{eqnarray}

\noindent
We have the same relations with $\delta_i \to \partial_i$ ( in fact
$\delta_i \equiv \left(\partial_i\right)^{\star}$  see after).
These relations have been already derived by Mohammedi in Ref. 
\cite{fsusy1d}.
Alternative derivation through matrix realization of the algebra has been
obtained in Ref. \cite{fr}. A third derivation using a commuting set of 
GGA and changing the statistics through a Klein transformation is detailed in
the Appendix.
 
Returning to our heterotic extension of the complex plane, we   can however
define an automorphism of the algebra exchanging 
($z,\theta_L$) and ($\bar z,\theta_R$).
The algebra defined in relations (6) and (10-11) is NEITHER stable   under   
complex 
conjugation, NOR under the permutation of the $\theta$'s indices 
(we note $\sigma$ this permutation).
However, it is stable under the composition of both
 
$$  (AB)^{\star \circ \sigma} =   A ^{\star \circ \sigma} 
B^{\star \circ \sigma}.$$
\noindent 
With such an automorphism, $(z,\theta_L,\partial_L,\delta_L)$ is mapped onto
$(\bar z,\theta_R, \delta_R,\partial_R)$ and {\it vice   versa}, so we see
that we have a connection between the right-handed and the left-handed part
of the action.
Stress that under this conjugation, $\partial_L$ is exchanged with
$\delta_R$.
 The next
point before the construction of the action, is to remark that $(\partial 
\theta^a)^\star
=\theta^a \delta$, where $\partial$ acts from the {\it right}
and $\delta$ from the
{\it left}, this can be seen directly on the matrix
realization of the algebra
\cite{fr} and from $\partial^\star = \delta$, $\theta^\star=\theta$.

If we set 
\begin{eqnarray}
D_L &=& \delta_L + q^2 \theta_L^2 \partial_z \nonumber \\
Q_L &=& \partial_L + q \theta_L^2 \partial_z,
\label{2.3}
\end{eqnarray}

\noindent 
respectively the covariant derivative and the FSUSY generator       
associated to the $z-$ modes. We obtain under the  $\star \circ \sigma$
conjugation
the  covariant derivative and the  FSUSY generator of the $\bar z -$modes
\begin{eqnarray}
D_R &=& \partial_R + q  \theta_R^2 \partial_{ \bar z} \nonumber \\
Q_R &=& \delta_R + q^2 \theta_R^2 \partial_{\bar z}.
\label{2.4}
\end{eqnarray}

\noindent
A direct calculation proves that $D_L^3=Q_L^3=-\partial_z$ and
$D_R^3=Q_R^3=-\partial_{\bar z}$, as in $1D$.

\noindent
Following Azc\'arraga and Macfarlane \cite{am}, $D_L$ (resp. $D_R$) acts from
the left (resp. the right). Introduce the fractional superfield,

\begin{eqnarray}
\Phi(z,\theta_L, \bar z ,\theta_R) &=&
X(z ,\bar z) + q^2 \theta_L \psi_{20}( z ,\bar z) +  
q^2 \theta_L^2 \psi_{10}( z ,\bar z) \nonumber \\
&+& q^2 \theta_R \psi_{02}( z ,\bar z) + \theta_L \theta_R \psi_{22}( z, \bar z)
+ q^2 \theta_L^2 \theta_R  \psi_{12}( z ,\bar z) \\
&+& q^2 \theta_R^2 \psi_{01}( z ,\bar z) + q^2 \theta_L \theta_R^2
 \psi_{21}( z ,\bar z) + \theta_L^2 \theta_R^2 \psi_{11}( z, \bar z). \nonumber
 \label{2.5}
 \end{eqnarray}

\noindent 
The  components  $\psi_{ab}$ ($X=\psi_{00}$) are of grade $a+b$ and 
satisfy, because of their grade,
 
\begin{eqnarray}
\theta_L \psi_{ab} &=& q^{a+b} \psi_{ab} \theta_L \\
\theta_R \psi_{ab} &=& q^{a+b} \psi_{ab} \theta_R. \nonumber
\label{2.6}
\end{eqnarray}
  
Now we are ready to build the $2D-$ action $S$. With similar arguments as those
used in $1D$, and with $D_L (D_R)$ acting from the left(right) we get 

\begin{eqnarray}
S &=& q \int dz d\bar z d \theta_R d \theta_L~
\left[ D_L \Phi(z,\theta_L, \bar z ,\theta_R) 
\Phi(z,\theta_L, \bar z ,\theta_R) D_R \right]  \nonumber \\
 &=& \int dz d\bar z ~ [\partial_z  X(z , \bar z) \partial_{\bar z} 
 X(z , \bar z)\nonumber \\ 
&&~~~~~~~~~~~~~- q\partial_z \psi_{02}(z, \bar z) \psi_{01}(z, \bar z)
+ q^{2}\partial_z \psi_{01}(z, \bar z) \psi_{02}(z, \bar z) \nonumber \\ 
&&~~~~~~~~~~~~~+ q\psi_{20}(z, \bar z) \partial_{\bar z } \psi_{10}(z, \bar z) 
 -q^2 \psi_{10}(z, \bar z) \partial_{\bar z } \psi_{20}(z, \bar z) \\ 
&&~~~~~~~~~~~~~-q \psi_{11}( z, \bar z) \psi_{22}( z, \bar z)   
 - q^2 \psi_{22}( z, \bar z) \psi_{11}( z, \bar z) \nonumber \\ 
&&~~~~~~~~~~~~~+\psi_{12}( z, \bar z) \psi_{21}( z, \bar z)
  +\psi_{21}( z, \bar z) \psi_{12}( z, \bar z) ~]. \nonumber
\label{2.7}
\end{eqnarray}  
  
First, note that this action is a grade $0$ number. Second, if we
choose $\psi_{ab}^\star =  \psi_{ba}$
and also with the appropriate
choice of the power of $q$, in the definition of $\Phi$, we ensure that the 
Lagrangian is real. 
Solving the equations of motion, we see that: 
  
 $\bullet$ $X$ admits a holomorphic and an antiholomorphic part;

 $\bullet$ $\psi_{10},\psi_{20}$ are holomorphic;
 
 $\bullet$  $\psi_{01},\psi_{02}$ are antiholomorphic;
 
 $\bullet$ $\psi_{12},\psi_{21},\psi_{11}, \psi_{22}$ are auxiliary fields that
  vanish on-shell. 
  
  \noindent  
En the $1D$ case, no modes expansion of the fields are allowed
(except in the path integral formalism where developments on the
eigenvectors can be used). However, in $2D$ (and upper dimensions), 
nothing can be said on the $q-mutation$ of the various fields, 
but only on the modes of their associated Laurent expansions 
(see (10-11)). We will come back to
this point further.
  
\noindent  
Finally, let us introduce $\epsilon_L$ and $\epsilon_R$ the parameters of the
FSUSY transformations. Utilizing 
  
  (1) the structure of the algebra, for the
  $L$ and $R$ handed sectors ($Q_L D_L = q^2 D_L Q_L $ and
  $Q_R D_R = q D_R Q_R$), and from the fact that the covariant derivative
  has to commute with the FSUSY transformations;
  
  (2) an ordering upon the variables consistent with the algebra (see (10))
  and the $\star\circ \sigma$ automorphism;
  
  \noindent 
we get the  following $q-$mutation relations

\begin{eqnarray}
\epsilon_L \epsilon_R& =& q \epsilon_R \epsilon_L \nonumber \\ 
\epsilon_L \theta_R &=& q \theta_R \epsilon_L \nonumber \\ 
\epsilon_L \theta_L &=& q^2 \theta_L \epsilon_L \\  
\epsilon_R \theta_L &=& q^2 \theta_L \epsilon_R \nonumber \\ 
\epsilon_R \theta_R &=& q \theta_R \epsilon_R, \nonumber
\label{2.8}
\end{eqnarray}
  
\noindent 
and the FSUSY transformations of the fields $\Phi$
 
\begin{equation}
\delta_\epsilon \Phi = \epsilon_L Q_L \Phi + \Phi Q_R \epsilon_R
\label{2.9}
\end{equation}
 
\noindent
or in components

\begin{eqnarray}
\delta_\epsilon X &=& q^2 \epsilon_L \psi_{20} + q \psi_{02} \epsilon_R 
\nonumber \\ 
\delta_\epsilon \psi_{20} &=& -q \epsilon_L \psi_{10} + q^2 \psi_{22} \epsilon_R
\nonumber \\ 
 \delta_\epsilon \psi_{10} &=& \epsilon_L \partial_z X +\psi_{12} \epsilon_R
\nonumber \\ 
\delta_\epsilon \psi_{02} &=& q^2 \epsilon_L \psi_{22}- q^2 \psi_{01} \epsilon_R
\\
\delta_\epsilon \psi_{01} &=& q^2 \epsilon_L \psi_{21} + \partial_{\bar z} X
\epsilon_R \nonumber \\ 
\delta_\epsilon \psi_{22} &=& - q \epsilon_L \psi_{12}-  \psi_{21} \epsilon_R
\nonumber \\ 
\delta_\epsilon \psi_{11} &=&   \epsilon_L \partial_z \psi_{01} +  
q ^2 \partial_{\bar z}  \psi_{10} \epsilon_R
\nonumber \\ 
\delta_\epsilon \psi_{12} &=&   \epsilon_L \partial_z \psi_{02} -
q\psi_{11} \epsilon_R \nonumber \\ 
\delta_\epsilon \psi_{21} &=&   -q\epsilon_L   \psi_{11} +
q  \partial_{\bar z} \psi_{20} \epsilon_R.
\nonumber
\label{2.10}
\end{eqnarray}
 
The form of the action (16) is legitimated by the fact that the 
component $\theta_L^2  \theta_R^2$ transforms as a total derivative under
FSUSY. Furthermore, the action is also invariant under conformal
transformations. 
  
\vskip .5truecm
\noindent
{\bf  IV.  The Green Functions and the Wick contraction.}\\
\vskip .5truecm
This section is devoted to the calculation of the Green functions associated
to the action (16). Here, we will focus our attention
on the holomorphic part of the action and we note $X,~\psi_1,~ \psi_2$
the basic fields. Two equivalent
calculations will be proposed: the path integral approach and the mode expansion
 one. The latter will be useful for the normal ordering prescription and
 the operator product expansion (OPE) of the algebra.
\vskip .3 truecm
\noindent
{\underbar{IV.1 The Path Integral Approach.}}
\vskip .3 truecm

\noindent
We want to calculate the partition function $Z$

\begin{equation}
Z[0]=\int {\cal D} \psi_2 {\cal D} \psi_1 \exp{ \left[ \int  dz d \bar z \left(
q \psi_2(z,\bar z)
\partial_{\bar z} \psi_1 (z,\bar z)- q^2 \psi_1(z,\bar z)
\partial_{\bar  z} \psi_2 (z,\bar z) \right) \right]}
\label{3.1}
\end{equation}

\noindent
Point out that the order of the path integration is opposite to the action
one, in order to avoid the unwanted $q$-factor.

In (20), $\psi_1$ and $\psi_2$ are defined in the complex plane.
In a discretization process, we just particularize the case where they
are $N$-component vectors. On the same footing, the kinetic operator
becomes a $N\times N$ matrix, noted $A$. So, we have to compute
\begin{equation}
Z[0]=\int (d \psi_2)^N (d \psi_1)^N \exp{\left( \psi_1 A \psi_2\right)}
\label{3.2}
\end{equation}
\noindent
It is known that  any bilinear form can be diagonalized by two different
transformations of determinant one, $\Delta = J A J'$. Using the property
upon the integration on
GGA variables $\int(d \theta)^N = (\det J)^{-2} \int (d [J\theta])^N$
 \cite{fr}
(this can be seen directly from $\int d \theta = {d^2 \over d \theta^2}$,
with an affine transformation) we get 

 \begin{equation}
Z[0]=\int (d \psi_2)^N (d \psi_1)^N \exp{\left( \psi_1 A \psi_2\right)} 
= det(A)^2
\label{3.3}
\end{equation} 

\noindent
So, we obtain 

\begin{eqnarray}
Z[0]&=&\int {\cal D} \psi_2 {\cal D} \psi_1 \exp{ \left( \int dz d \bar z
   \pmatrix{\psi_1&\psi_2}
\pmatrix{0&-q^2\partial_{\bar z}\cr
         q\partial_{\bar z}&0} \pmatrix { \psi_1 \cr
        \psi_2} \right) }\\
        &=&det  \pmatrix{0&-q^2\partial_{\bar z}\cr
         q\partial_{\bar z}&0}^2\nonumber
\label{3.4}
\end{eqnarray} 
 
\noindent
Of course, the measure of integration has been defined in an appropriate way,
such that the path integral (20) is just equal to $(\det A)^2$,
in other words, each integral over $\psi_1$ and $\psi_2$ has been
normalized by a $1 \over \sqrt{2} $ term and some phase factors.
This result has been already obtained by  Matheus-Valle {\it et al} in Ref. 
\cite{fsusy2d}, and can be obviously extended for a GGA of any order
($\theta^n=0$).

The two points Green function can be derived using the usual procedure, where
two GGA sources are introduced (see for example  Matheus-Valle 
{\it et al} in Ref.\cite{fsusy2d}). Here, we propose 
an alternative calculation with respect to the kinetic operator $A(z-w)$.
The action can be rewritten in an equivalent way
$$S=\int d^2 z d^2 w \pmatrix{\psi_1(z)&\psi_2(z)} A(z-w)
\pmatrix { \psi_1(w) \cr
\psi_2(w)}$$
\noindent      
with,
$$A(z-w)=\pmatrix{0&-q^2\partial_{\bar z}\cr
         q\partial_{\bar z}&0} \times \delta(z-w)\delta({\bar z}-{\bar w})$$
\noindent
The propagator is then
  
\begin{eqnarray}  
<\pmatrix { \psi_1(z) \cr \psi_2(z)} \pmatrix{\psi_1(w)&\psi_2(w)}>
 &=&{\delta \over \delta A(z-w)}  Z[0] \nonumber \\                     
 &=& \pmatrix{0&-q\cr
              q^2&0} { 1 \over z-w}.  
\label{3.5}
\end{eqnarray}

\noindent
In this derivation, to avoid the unwanted $2$ factor coming from the derivation
of $(\det A)^2$, each fields is normalized with a $1 \over \sqrt{2} $ factor,
as for the measure of integration.
From (24) and from the well-known result on the
propagator of scalar fields in $2D$, we can deduce
the none-vanishing propagators 

\begin{eqnarray}
< X(z) X(w) > &=& (-\partial_z \partial_{\bar z})^{-1}
=-\ln(z-w) \nonumber \\
< \psi_1(z) \psi_2(w) > &=& { q^2 \over z-w} \\
< \psi_2(z) \psi_1(w) > &=& { -q  \over z-w}. \nonumber
\label{3.6}
\end{eqnarray}
From the propagator of $\psi_1$ and  $\psi_2$, it seems that they fulfill
braiding properties, although they do not. This discrepancy will be explain 
further, in the next sub-section.

\vskip .3 truecm
\noindent
{\underbar{IV.2 The Modes Expansion}}
\vskip .3 truecm

\noindent
First of all, as we will justify in the next section, note that the
fields $\psi_1$ and $\psi_2$ are respectively of conformal weight
$2/3$ and $1/3$. Following the standard convention in  
Conformal Field Theory (CFT), their modes expansion can be expressed
     
\begin{eqnarray}
\psi_1(z)&=&\sum\limits_{r_1 \in \ZZ + {a \over 3}}
 \psi_{1,r_1}z^{-r_1-{2\over 3}}\\
\psi_2(z)&=&\sum\limits_{r_2 \in \ZZ + {2a \over 3}}
 \psi_{2,r_2}z^{-r_2-{1\over 3}}
\nonumber
\label{3.7}
\end{eqnarray}
By analogy with the string case, according to the value of $a$ ($a=0,1,2$),
we will have different boundaries conditions 
($z \rightarrow \exp{(2 i \pi)} z$) for the $\psi_i$ fields.

$\bullet$ For $a=0$, $\psi_1$ picks up a $q$-phase factor and  
$\psi_2$ a $q^2$ one.  
  
$\bullet$ For $a=1$, $\psi_1$ and $\psi_2$ remain unaffected.
   
$\bullet$ For $a=2$, $\psi_1$ picks up a $q^2$-phase factor and 
$\psi_2$ a $q$ one.

In these three situations, the Lagrangian remains, of course, unaffected.
Point  out that these sectors are adapted extension of the Ramond
and Neveu-Schwarz \cite{gsw} ones.
Using the modes of the $\psi$ fields, we can identify some of them with
the $\theta_i$ or $\partial_i$ . If one mode of $\psi_1$ is associated
to $\theta$, the corresponding mode of $\psi_2$ has to be associated to
$\partial$.  Of course, it depends on the definition of the vacuum.
Setting the following convention, associated to special choice of the vacuum,
we obtain: 

\begin{eqnarray}
\psi_{1,r_1} | 0 >&=& 0,~~~~ r_1 > 0 \\
\psi_{2,r_2} | 0 >&=& 0,~~~~ r_2 > 0, \nonumber
\label{3.8}
\end{eqnarray}

\noindent
and the  $q-$mutation relations (which corresponds to the 
identification $\psi_{1,r} \equiv \theta_i$,  $\psi_{2,r} \equiv 
\partial_{\theta_j}$ and the algebra 
(10-11)). Through this identification,
we can in principle write all the $q-$mutation relations among the
various modes. However, for our purpose we {\it only} need to know the
$q-$mutations when the indices do not have same signs.

\begin{eqnarray}
&&\psi_{2,r_2} \psi_{1,r_1} = q^2 \psi_{1,r_1} \psi_{2,r_2},~~~~~ r_2 <0, ~~ 
r_1>0,~~~~r_2 \ne - r_1, \nonumber \\
&&\psi_{2,r_2} \psi_{1,r_1} = q \psi_{1,r_1} \psi_{2,r_2},~~~~~~ r_2 >0,~~
r_1<0,~~~~r_2 \ne - r_1, \nonumber \\ 
&&\psi_{2,r} \psi_{1,-r} -q \psi_{1,-r} \psi_{2,r} = -q,~~~~~ r>0, \\
&&\psi_{2,-r} \psi_{1,r} -q^2 \psi_{1,r} \psi_{2,-r} = -q,~~~~ r> 0, \nonumber 
\\
&&\psi_{2,r} \psi_{2,s} = q \psi_{2,s} \psi_{2,r},~~~~ r <0,~~~~ s>0, 
\nonumber \\
&&\psi_{1,r} \psi_{1,s} = q \psi_{1,s} \psi_{1,r},~~~~ r <0,~~~~ s>0, 
\nonumber  
\label{3.9}
\end{eqnarray}

\noindent
those relations are very close to those obtained in one-dimension where
quantization {\it \`a la Dirac} where used \cite{fr}.
Notice that the derivative of $\theta$ is obtained using a change in the sign
(see the third and fourth equations of (28)), in accordance with
(26). From the choice of the
vacuum (27) there is one and only one correspondence between the modes
of $\psi_1,\psi_2$ and the generators of the algebra  (10-11). It has to be 
emphasized, using (28), that nothing can be said on the 
$q-$mutation on the various fields but ONLY on their modes. 
In other words, nothing can be said on
the symmetry of the wave function, but only on the states in the Hilbert space.
From the definition of the vacuum, it becomes now easy
to derive the same propagator as before for the $\psi$'s fields. In the 
derivation of (24), using the modes expansion, ones see immediately the 
braiding property of the propagator because of the definition of
the vacuum. This property is however lost is the general case: when we 
calculate $<\psi_1(z_1) \psi_2(z_2) {\cal O}(z_3)>$ for an arbitrary
operator ${\cal O}$.

To define the normal ordering, we proceed as usual putting to the right the
creation operators and using explicitly the algebraic structure.
From the identification

\begin{eqnarray}
 \psi_{1,r<0} &\sim& a_{-r}^+   \nonumber \\
 \psi_{2,r>0} &\sim& a_r \nonumber \\
 \psi_{2,r<0} &\sim& b_{-r}^+     \\
 \psi_{1,r>0} &\sim& b_r   \nonumber 
 \label{3.10}
\end{eqnarray}

\noindent
the connection between the $\psi$ variable and the $q-$oscillators can be
found in Ref. \cite{q-osc}. It has to be stressed that such an identification
enable an explicit construction of the modes of the fields (so the fields 
themselves) and their basic $q-$mutation relations,
only from the coherence of the algebra (10-11).

\noindent 
Stress that, if nothing can be said on the $q-$mutation between the fields  
$\psi_1$ and $\psi_2$, it is no more the case through a normal ordering.
This is a consequence of the peculiar structure of the algebra (28).   
Indeed, we have 

\begin{equation}
:\psi_1(z) \psi_2(z): = q^2 :  \psi_2(z) \psi_1(z):.
\label{3.11}
\end{equation}

\noindent
To get this equation we have used the definition of the vacuum,
the $q-$mutation
among the modes and a regularization of a term like $a_r^+b_s^+$.  

\noindent
From this normal ordering prescription, if one wants to determine a $4-$points
Green function, using the Wick theorem, one is faced immediately with an
ambiguity. In fact, if we calculate
naively  for instance : $<:\psi_1(z) \psi_2(z):
:\psi_1(w) \psi_2(w):>$ we have two possibilities leading
to different results :

$\bullet$ we can do first the contraction of $\psi_2(z)$ with $\psi_1(w)$, and
then $\psi_1(z)$ with $\psi_2(w)$;

$\bullet$ we can $q-$mute the $\psi$'s inside the two normal ordering and then
do the contraction of $\psi_1(z)$ with $\psi_2(w)$ and  
$\psi_2(z)$ with $\psi_1(w)$.

\noindent
The result of those two different calculus will differ by a factor $q$ !
This result never appears with fermions or bosons because
the signs will compensate. This problem can be solved as follow:
 
Arguing that if $ : AB : ~=~ q : BA :$, we do not have usually, after a
transformation, $ \delta (: AB :) ~=~ q \delta (: BA :) $. In order
to be coherent, we have to impose for the variation of
$: AB: $ the natural substitution 
${1\over 2} ( \delta (: AB :) ~+~q \delta (: BA :) )$. 
Paying attention that $T$ and $G$ are respectively the generators
of the conformal and fractional supersymmetry transformations, we
have to substitute in the two points Green functions  involving
$T$ or  $G$ expressions analogous to 
$${1\over 2}(<:\psi_1(z) \psi_2(z):~:\psi_1(w) \psi_2(w):>~+~
q<:\psi_2(z) \psi_1(z):~:\psi_2(w) \psi_1(w):>).$$
This procedure can be extented
for $N-$ points Green function using similar permutations.

Finally, before closing this section, we want to mention that, one can build
propagator involving fractional power of $(z-w)$, using other 
algebraic structure than (10-11) or GGA. 
This is the essential difference of our result with respect to Saidi 
{\it et al}. This can be understood as follow\footnote{This point has been 
mentioned to us by D.~ Bernard.}:
$${1\over {(z-w)^{\Delta}}}={1\over {z^\Delta}} \sum\limits_{n\geq 0}
a_n(\Delta)
({w\over z})^{\Delta}~~~, |z| > |w|.$$

If we change explicitly the $q-$mutations relations of $\psi_{1,r}$ with 
$\psi_{2,-r}$ by introducing on the RHS the appropriate numbers $a_n(\Delta)$  
in Eq. (28), one can get $$< \psi_1(z) \psi_2(w) > 
\sim { 1 \over (z-w)^\Delta}.$$ On the level of path integration, 
new rules have to be derived, substituting the result obtained in Eq. (22).

\vskip .3 truecm
\noindent
{\bf V. Current Algebra within Fractional Supersymmetry.}
\vskip .3 truecm
In this algebra, we have three different fields $X(z)$, $\psi_1(z)$
and $\psi_2(z)$ on which act two symmetries: the conformal and FSUSY
transformations. The former will be generated by the stress momentum
tensor $T(z)$ and the latter by the fractional supercurrent $G(z)$.
\noindent
In this section, as in the previous one, we consider only the holomorphic
part.
\vskip .5cm
\noindent
\underbar{ V.1 The stress momentum tensor.}
\vskip .5cm
\noindent
There are three ways to derive the stress momentum tensor $T(z)$.

1) First, coupling the different fields $X,\psi_1, \psi_2$ to the gravitational
field, using the appropriate covariant derivative
and invoking the standard general
relativity results (\cite{gsw}, p.62).
\noindent 

2) The second, using Polyakov's results: coupling  the fields in a
 non-conformally flat
metric and performing an adapted transformation upon the variables 
 (\cite{p}, p.236).

3) The third one is the most tractable for our purpose. From dimensional
arguments, we deduce the conformal weights of the various fields. Using
the definition of $T$ as well as the transformation property of conformal
field with conformal weight $h$, we get the expression of $T$. 
Let us detail this approach.

\noindent
$X$ is a conformal weight $0$, so is $\Phi$, the fractional superfield.
$D_{L}^3=-\partial_{z}$, so $D_{L}$, $\theta$ are of conformal weight
$1/3$ and $-1/3$ respectively. It means that $\psi_1$ and $\psi_2$
are of conformal weight $2/3$ and $1/3$. So,
\begin{equation}
T(z)=-{1\over 2}~:\partial_z X(z) \partial_z X(z):~+
{2\over 3} q^2 ~:\psi_1(z) \partial_z \psi_2(z):~-
{1\over 3} q^2 ~:\partial_z \psi_1(z)\psi_2(z) :
\label{4.1}
\end{equation}
\noindent
Note that $T$ is of grade $0$.
Using the Wick theorem as well as the basic propagators, one can check
\begin{eqnarray}
T(z) X(w)&=&{\partial_w X(w) \over (z-w)}+\dots,~~~|z| > |w|\nonumber \\
T(z) \psi_1(w)&=&{{2\over 3} \psi_1(w) \over (z-w)^2} +
{\partial_w \psi_1(w) \over (z-w)} +\dots,~~~|z| > |w|\\
T(z) \psi_2(w)&=&{{1\over 3} \psi_2(w) \over (z-w)^2} +
{\partial_w \psi_2(w) \over (z-w)}+\dots,~~~ |z| > |w|\nonumber
\label{4.2}
\end{eqnarray}
as it should be. The $\dots$ represents the regular part of the O. P. E.'s 

\noindent
To prove the consistency of the algebra, the next point is to check the 
action of $T$ on itself. After a little algebra, paying attention
on the Wick contraction for $4-$points Green function (see sect. IV.2),
we have

\begin{equation}
T(z) T(w) = {{1\over 2} (1+{2\over 3})   \over (z-w)^4} +
{2 T(w) \over (z-w)^2} +
    {\partial_w T(w) \over (z-w)} + \dots,~~~ |z| > |w|
\label{4.3}         
\end{equation}

\noindent
In this O. P. E. the anomaly has two origins: one arising from the
scalar field,
as usual, and one from  the $\psi$'s fields. We will come back to
this point in the conclusion and outlooks.
It has to be stressed that the action (16) leads naturally to a 
description of the fields with rational conformal weight ( in this
simplest case
$1/3,2/3$). So, the FSUSY transformation is a symmetry which connects
states of spin $0,1/3,2/3$, generalizing, in that sense,
 the notion of supersymmetry.
Before concluding this sub-section, we can say few words on the stress-momentum
tensor.
It is known that in $2D$, in addition to the invariance of the complex plane,
one has the Weyl invariance acting on the metric: 
$g \rightarrow e^{\phi(z, \bar z) }g$.
With such a symmetry, plus the diffeormorphism 
$z \rightarrow f(z,\bar z),~~\bar z \rightarrow \bar f(z,\bar z)$
we can globally eliminate  the gravitation. In this special gauge,
remains just the conformal symmetry which just transforms the metric up
to a scale factor. So, the derivative has to be substituted by the appropriate
covariant derivative ( see {\it e.g.} \cite{gsw}, p.126).
In this case we get

\begin{eqnarray}
\nabla_z \psi_1(z)& =& \partial_z \psi_1(z) -{2 \over 3} \partial_{ z} 
\phi(z),  \nonumber \\
\nabla_z \psi_2(z)& =& \partial_z \psi_2(z) -{1 \over 3} \partial_{ z} 
\phi(z),  \nonumber
\end{eqnarray}

\noindent
with such a definition the stress-momentum tensor can be expressed with
the normal or covariant derivatives because the Christoffel's symbols
cancel.

 \vskip .5cm
\noindent
\underbar{ V.2 The Fractional Supercurrent.}
\vskip .5cm
\noindent
Using the results of Durand in Ref. \cite{fsusy1d}, as well as the  
$2D-$FSUSY transformations (19) we get

\begin{equation}
G(z) = -q^2(~ :\partial_z X(z)~ \psi_2(z):~+{1 \over 2} ~:\psi_1^2(z):~)
\label{4.4}  
\end{equation} 

\noindent
Along the same lines, as for the action of $T$ on the fields, we can reproduce
the FSUSY transformations on the fields

\begin{eqnarray}
G(z) X(w) &=& {q^2 \psi_2(w) \over (z-w)}+ \dots,~~~|z| > |w| \nonumber \\
G(z) \psi_1(w) &=& {\partial_w X(w) \over (z-w)}+ \dots,~~~|z| > |w|   \\
G(z) \psi_2(w) &=& {-q \psi_1(w) \over (z-w)}+ \dots,~~~|z| > |w| \nonumber
\label{4.5}  
\end{eqnarray} 

\noindent
It has to be stressed that the action of the supercurrent on the fields
gives the same transformation properties as in relations (19), as it should be.
\noindent
Now, it remains to check the closure of the algebra .
We can calculate successively 

\begin{eqnarray}
T(z) G(w) &=& {{4 \over 3} G(w) \over (z-w)^2} + 
{\partial_w G(w) \over (z-w)} + \dots,~~~|z| > |w| \\
 G(z) G(w) &=&  {-q :~ \psi^2_2(w) ~ : \over (z-w)^2} +
{ \tilde  G(w) \over (z-w)} + \dots,~~~|z| > |w|, \nonumber
\label{4.6}  
\end{eqnarray}

\noindent
with $\tilde G(z) = (1-q^2) ~:~ \partial_z X(z) \psi_1(z)~:~ 
-q~:~ \partial_z \psi_2(z) \psi_2(z)  ~:~ $. 
The first of these relation just tells us that $G$ is a conformal field
of conformal weight $4/3$.
Now comes the question about
$\tilde G(z)$ : is it a generator of a symmetry ? In other words can we
find, in addition to the conformal and the FSUSY transformations,
other symmetries of the complex plane.  
 
Looking to the algebraic structure, one can see that
those two symmetries are the only ones. In fact we cannot
find symmetry with generators of grade $1$ (the conformal/FSUSY transformations
is generated by a grade $0$/$2$ operator) {\it i.e} acting only on
$\theta_L^2$ and leaving  $\theta_L$ unchanged. This result has also
been proved on the level on the lagrangian by Matheus-Valle
{\it et al} in \cite{fsusy2d}.
Finally, using Durand's result derived in
\cite{fsusy1d}, {\it only} $Q$ and $Q^3$ are generators of symmetry
because they are the only ones that fulfill the Leibnitz rules.
We will come back to this point in the next section.
So, the new feature of these algebra, is that it will close under 
ternary and not bilinear identities. This is a reminiscence of the fact
that the basic algebra (generated by $Q_L$ and $\partial_z$, see sect. III ) 
is not Lie or graded Lie algebra. Finally we can check the closure of 
the algebra.

\begin{eqnarray}
(G(z) \tilde G(w) + \tilde G(z) G(w))& =&
{(-2 +{q \over 2}) \over (z-w)^3} - {~:~ \psi_2(w) \psi_1(w)~:~ \over (z-w)^2} 
\\
&-&{ 6 T(w) + ~:~\psi_1(w) \partial_w \psi_2(w)~:~ +  
~:~\psi_2(w) \partial_w \psi_1(w) ~: 
\over (z-w)} \nonumber
\label{4.7}  
\end{eqnarray}

\noindent
In these relations, we have taken a symmetric product for $G$ and $\tilde G$,
in order to ensure the associativity of the algebra. To get this O.P.E.
with have used $~:~\partial_z \psi_2 \psi_2 ~:~ =
q^2 ~:~\psi_2 \partial_z\psi_2 ~:~ $ arising from relation (28).
The algebra has the
special feature to close under ternary relations $G G G $, and with 
quadratic dependence on the fields.
This algebra can be compared with the fractional superconformal algebra
introduced in Ref. \cite{fconf} which  is also generated, in addition to
the stress momentum tensor, by a current of conformal weight $4/3$.
These two extensions of the Virasoro algebra are different. 
The fractional superconformal algebra closes with rational 
power of $z-w$, leading to non-local algebras because cuts are involved.
The one we propose, closes only with {\it integer} power of $z-w$
but involves cubic relations instead of quadratic ones. 
In a similar way, we can mention that this feature is not
specific to our model and already appears  in the framework
of the $W_n$ algebra where polynomial dependence  
of the generators are involved to close the algebra \cite{w}.
However, invoking
the remark done at  the end of the  previous section, 
by the appropriate substitution 
in Eq. (28), one should obtain the O.P.E. of the fractional
superconformal algebra, {\it i.e.} with fractional powers of $(z-w)$. Of 
course, correlatively, the spin of the $\psi_1$ and $\psi_2$ fields
change leading to different families of integrable models.

\eject 
\noindent
 {\bf VI. Beyond 2d FSUSY. }
\vskip .5cm
\noindent
In our paper, we have considered only two kinds of symmetries, one 
coming from the $2D$ conformal invariance and the second from the
$2D$ FSUSY. We have argueed that those symmetries
were the only one to be considered in  {\it our construction}. 
In this section, we are going to justify this point using algebraic
arguments, and we will gauge the symmetries. 

The fractional superfield $\Phi$ is defined
in some appropriate representation
of the fractional superspace $(z, \theta_L, \bar {z}, \theta_R)$. So,
the symmetries of $(z, \theta_L, \bar {z}, \theta_R)$ acting on 
$\Phi$ are built up with the differential operators of the fractional
superspace ($\partial_z, \partial_{\theta_L}, \delta_{\theta_L}$).
For the sake of simplicity, we will consider in the following
only the $L-$movers. 

1) From the basic differential operators
$\partial_z, \partial_{\theta_L}, \delta_{\theta_L}$,
one can build {\it a priori} grade $0, 1, 2$ operators. 
In this construction, because $\partial_{\theta_L}$
and $\delta_{\theta_L}$ are of grade $2$, only grade $2$ and $0$
operators can be built using first order differential operators 
( for example, $Q$ and $\partial_z$, the generators of FSUSY and conformal
transformations respectively belong to this category).

2) Arguing, that respectively $\Phi$ and $D \Phi$ have to transform 
in the same
way, the covariant derivative has to commute with the generators of
the symmetries. So, using the algebraic structure of GGA, the only
allowed solutions are  $\partial_z,Q$ and $ Q^2$. 

3) The third point, in this argumentation, implies that the product of
two fractional superfields has to be also a fractional superfield.
As $Q^2$ does not verify the Leibnitz rule, the once symmetries
retained, in order to build invariant Lagrangian, 
are the FSUSY and the conformal transformations.

4) The remaining conserved Noether currents are $T$ and $G$.
In consequence, the would be conserved current associated to $Q^2$, 
something like $\tilde G$ defined in eq. (36), is not conserved
and does not belong to the algebra. So, using the two generators
$T$ and $G$, we close the underlying basic symmetry using cubic relations.
 
Indeed, all these points can be extended for any $F$, $F$ being the
order of the FSUSY transformations. 

Clearly, all these  assertions are relevant within the framework of our
kind of construction. Extension of the underlying symmetries in other
approaches could in principle considered.
This leaves the potentiality to add
$Q^2$ as a generator of the associated symmetry. This peculiar property has 
been exploited in the second paper of \cite{ssz} where a conserved spin $5/3$ 
current  were introduced in addition to a spin $4/3$ current.
Following \cite{ssz}  the algebra $\theta^3=0$ 
can be represented in a linear way introducing two Grassmann variables
$\theta_1$ and $\theta_2$ satisfying
$$ (\theta_1)^2  =\theta_2,~~ \theta_1 \theta_2 + \theta_2 \theta_1=0.$$
With this associated representation of the algebra, two generators 
(having a linear dependence in the previous variables and their derivatives)
can be introduced. 
However, the possible representation consistent with the algebra (10-12) 
is not at first glance obvious and needs further investigations.

In addition to this discussion, we want discuss the basic points
that lead, from the action defined in eq.(16) and invariant under global
transformations, to an action invariant under Gauge symmetries. 
Of course, the full Lagrangian will not be exhibited, but only 
the relevant points dictated by the Noether procedure introducing Gauge
fields that couple with their associated conserved current. The 
determination of the full invariant Lagrangian goes beyond the scope of 
this paper.

The $2D$ diffemorphism (which contains
the conformal transformations as a subgroup) are controlled by a metric
or a ``zweibein''. Similarly, the local FSUSY, {\it i.e.} the fractional
supergravity (FSUGRA) can be controlled by a Gauge field analogous to the
gravitino in supergravity, we have named fractino by analogy
\cite{fr}. 

Due to the non-linearity of the algebra, the existence of one
or two fractino(i) is still an open problem. 
In the first situation, we
add to the Lagrangian, using Noether procedure, a term like
$$\chi_1 G+\ldots ~~,\eqno(38)$$ 
where $\chi_1$ defines a fractino and $G$ the FSUSY current. 
Their  spin are $4/3$, $-4/3$ respectively. In the second situation,
the general Lagrangian has to be completed by an additive term including
two fractini $\chi_1$ and $\chi_2$:
$$\chi_1 G+\chi_2 \tilde{ G}+\ldots\eqno(39)$$ 
Let us point out that $\tilde G$ is the transformed field of $G$ under
FSUSY transformations: $\delta_{\epsilon} G= \epsilon \tilde{ G}$.
Arguing that $\epsilon$ is a spin $-1/3$ field, $\tilde G$ and
$\chi_2$ are of spin $5/3$ and $-5/3$ .
The presence of $\tilde G$ in (39)
can be seen as a reminiscence of the  peculiar   structure of the algebra.
Due to the fact that $Q$ closes under a cubic power instead of
a quadratic one, we might have two gauge fields instead of one
as in the framework of  Lie or super-Lie algebras. This second gauge field
$\chi_2$ is coupled to $\tilde G$, the would be conserved current of 
$Q^2$ (see point (4) of this section).

The various states of the spin for 
$\chi_1$, $\chi_2$ are respectively $-4/3, -2/3,2/3,4/3$ and
$-5/3, -1/3,1/3,5/3$ when the holomorphic and anti-holomorphic part
of the action have been considered. This peculiar form of the projection
can be explained as follow: $\chi_1$ is a vector-spin $1/3$ field, 
$\chi_1=\chi_{\pm 1, \pm {1 \over 3}}$ and in an analogous manner
$\chi_2=\chi_{\pm 1, \pm {2 \over 3}}$.

Using the Noether theorem, it is known that the fractino field $\chi_1$
has to transform like:
$$\delta_{\epsilon} \chi_1 \sim \partial_x \epsilon.$$
where $x$ stands respectively for $z$ or $\bar z$ according to
the value of the spin  of the fractino $\chi_1$.
Those results goes along the same line as in supersting theory for the 
transformation law of the gravitino see ( \cite{gsw}, p. 233).
With the same arguments as those employed in $2D$ SUGRA
and in connection with eq. (19),
among the ``zweibein'', $\chi_1$ and $\chi_2$ (which belong to the same
fractional superfields),  only the transformation of $\chi_1$ involves
derivatives.

In our construction, we can conclude that in addition to 
the conformal ghosts associated to the conformal 
transformations of the ``zweibein'', only the FSUGRA ghosts associated
to the $\chi_1$ transformation has to be considered.
This is because the {\it only} symmetries are generated by $\partial_z$ 
and $Q$.  We will come back to this point in the conclusion.

 \vskip .5cm
\noindent
{\bf VII. Conclusion and outlooks.}
\vskip .5cm
\noindent
We have obtained new structures extending the Virasoro algebra by considering
a generalization of conformal symmetry. This symmetry, as we have seen,
is a transformation which connects states of fractional spin. It has to be
stressed that those algebras are not constructed  from Lie or graded-Lie
algebra; meaning that they do not close via quadratic relations. 
Consequently, we have obtained conformal field theory which does not belong 
to the well- known model (fractional superconformal invariance)
 where fractional spin are involved \cite{fconf}.
 
The  conformal dimension of our CFT is $5/3$.
Taking into account this peculiar situation with $c=5/3$, 
new $2D-$ integrable 
models could be described using $2D-$ fractional supersymmetry.
Various extensions can be derived in this formalism. First of all,
we can modify the $q-$ mutation relations (28) leading to fractional
two points functions and to conformal fields with other conformal 
weight. Clearly, their associated central charge will change.
Secondly, instead of considering only a scalar superfield, we can
introduce a fractional superfield of conformal weight $h$.
Thirdly, it has been pointed out in \cite{ssz} that representations
of the FSUSY algebra can be obtained from the periodic representation of the 
quantum group $U_q(sl(2))$. Finally, one can introduce interactions,
including a superpotential
[N. Debergh in \cite{fsusy1d}], 
or a coupling between superfields of different conformal weight
[Colatto {\it et al} in \cite{fsusy1d}].

In the context of string, if FSUSY is the symmetry of the worldsheet, 
this symmetry could be used in order to build 
solutions with relevant phenomenology (appropriate gauge group,
three families of massless  quarks and leptons, space-time supersymmetry etc.).
It remains, of course,  an open question.
In that direction, we can easily calculate the critical dimension.
The conformal anomaly has two origins: one coming from the space-time
degrees of freedom($c_{X,\psi_1,\psi_2}=5/3 D$, $5/3$ for each dimension),
and the other
coming from the ghost-part of the action.  As in string theory,
the conformal part is $c_{b-c} = -26$ \cite{ps,gsw}. 
For the FSUSY part of the ghosts there is no need to
know the specific part of the  FSUSY-ghosts. From the conformal weight 
of  $\epsilon_L$ (the parameter of the FSUSY transformations) and the 
transformations properties of GGA we get (with  $J$, something like 
$\partial_{\bar z}$, the operator  of the transformation on the gauge 
field which controls the local FSUSY,  the fractino $\chi_1$). 

$$
S_{FSUSY} = \det J^{-2} = \int {\cal D} \beta {\cal D} \gamma 
{\cal D} \beta ' {\cal D} \gamma ' \exp{\left[S_{\beta, \gamma} 
+ S_{\beta ', \gamma '} \right]},
$$

\noindent
where $\gamma, \gamma '$ are two commuting ghosts of conformal weight $-1/3$
and  $\beta, \beta '$ two commuting ghosts of conformal weight $4/3$.
Following the results obtained by Polyakov, the stress-momentum is known
as well as  the contribution to  anomaly $c_{\beta, \gamma}=2\times 22/3$
(\cite{p}, p. 238). The critical dimension is then $D={34 \over 5}$. 
 
If one builds a theory with $Q^2$ as a additional generator, one needs another
pair of commuting ghosts of conformal weight $-2/3,5/3$. Following the 
same procedure as before, one can show  that their contribution to the anomaly
is $2 \times {23 \over 3}$, leading to a negative (!) critical dimension. \
So, in the context of string theory, the approach with {\it only} $Q$ as a 
generator should be more appropriate. Of course, the critical dimension 
$D= {34 \over 5}$ is meaningless, but for another $F$, appropriate  
integer dimension can eventually be reached.

\vskip .5cm
\noindent
{\bf Acknowledgments.}
\vskip .5cm
It is a pleasure to thank D.~ Bernard , N.~ Rivier, M.~ Rosso and
J.B.~ Zuber for useful discussions and remarks. 
\eject         
\noindent
 {\bf Appendix }
\vskip .5cm
\noindent
The purpose of this appendix is to construct explicitly the Klein 
transformation
adap\-ted to $q-$mutating numbers ($q=\exp{({2 i \pi \over 3})}$).
Consider a set $\xi_i,\partial_{\xi_i}$, $i=1 \dots p$ of commuting variables
satisfying the following relations.  

$$\partial_{\xi_i} \xi_i -q \xi_i\partial_{\xi_i} =1 \eqno(A.1).$$

\noindent
For the sake of simplicity, we develop the method only
for the derivative $\partial$; the case for the second derivative
$\delta$ is totally similar. 
\noindent
First, let us   introduce a number operator $N_i$ 

$$N_i =  \xi_i \partial_{\xi_i} +{(1-q)^2\over (1-q^2)}
 (\xi_i)^2 (\partial_{\xi_i})^2
\eqno(A.2),$$

\noindent
fulfilling the commutation relations 

$$ [ N_i, \xi_i ] =  \xi_i \eqno(A.3)$$
$$ [ N_i, \partial_{\xi_i} ] = - \partial_{\xi_i} \eqno(A.4).$$

\noindent
A direct calculation shows then that

$$   \xi_i q^{N_i} = q^{N_i+1} \xi_i =q  q^{N_i} \xi_i \eqno(A.5)$$
$$  \partial_{\xi_i} q^{N_i} = q^{N_i-1} \partial_{\xi_i} =
q^{-1}  q^{N_i} \partial_{\xi_i} \eqno(A.6).$$

\noindent
Introducing
$$ \theta_i = \xi_i \prod\limits_{j > i} q^{N_j} \eqno(A.7)$$
$$ \partial_i = \partial_{\xi_i} \prod\limits_{j > i} q^{-N_j} 
\eqno(A.8),$$

\noindent
one can check that the $\theta,\partial$'s satisfy the correct algebra
(10-11). So, we have built explicitly a cocycle, expressed in
terms of the basic fields, and allowing a change in the statistics {\it i.e.}
substituting commuting variables to $q-$muting ones. This is the principle
of the Klein transformation. Obviously this  can be extended in a similar 
way for any type of GGA ($\theta^n=0$). Notice that   
a number operator has already be introduced by Durand in \cite{fsusy1d}.
All these results can be easily found in the faithful matrix 
representation of Ref. \cite{fr}.

\eject

\end{document}